# Gas-like adhesion of two-dimensional materials onto solid surfaces

Zhengrong Guo[1,*], Tienchong Chang[1,2], Xingming Guo[1] and Huajian Gao[3]

[1]*Shanghai Institute of Applied Mathematics and Mechanics, Shanghai Key Laboratory of Mechanics in Energy Engineering, Shanghai University, Shanghai 200072, People's Republic of China*
[2]*State Key Laboratory of Ocean Engineering, School of Naval Architecture, Ocean and Civil Engineering, Shanghai Jiao Tong University, Shanghai 200240, People's Republic of China*
[3]*School of Engineering, Brown University, Providence, Rhode Island 02912, USA*

We conduct both atomistic simulations and theoretical modeling to show that there exists a reversible energy conversion between heat and mechanical work in the attachment/detachment of a two-dimensional material on/off a solid surface, indicating that two-dimensional materials adhesion is fundamentally like gas adsorption rather than solid adhesion. We reveal that the underlying mechanism of this intriguing gas-like adhesion is the entropy difference between the freestanding and adhered states of two-dimensional materials. Both the theoretical model and atomistic simulations predict that adhesion induced entropy difference increases with increasing adhesion energy and decreasing equilibrium binding distance. Our finding provides a fundamental understanding of the adhesion of 2D materials, which is important for designing two-dimensional materials based devices and may have general implications for nanoscale efficient energy conversion.

The discovery of graphene and its great success in many scientific areas have motivated continuous efforts to synthetize graphene-like materials, and consequently given birth to a new class of materials known as two-dimensional (2D) materials [1-3]. Owing to the extremely large surface of 2D materials, adhesion not only dramatically alters their inherent material properties, but also dominantly influences their synthesis [3], transfer [4] and device integration [5]. Indeed, the last decade has witnessed a continuously increasing interest on the adhesion of 2D materials from both scientific and technologic research fields [6-14]. So far, adhesion of 2D materials has been widely treated as mechanical contact between solid-like materials. In contrast to this mechanical contact perspective, a few recently found phenomena closely related to the adhesion of 2D materials show a strong temperature dependence [15-17]. For example, the intrawall adhesion can collapse a large-diameter carbon nanotube into a flat configuration, while at some higher temperature the collapsed tube can restore its cylindrical shape [15]. These phenomena imply a more complex picture of 2D materials adhesion that is fundamentally different from bulk solids adhesion.

---
[*] guozhengrong@shu.edu.cn

In this letter, we shown via molecular dynamic simulations that the adhesion of 2D materials to other surfaces is significantly influenced by thermal fluctuations. There in fact exists heat release/absorption in the attachment/detachment of a 2D material onto/from a substrate, indicating that the adhesion of 2D materials is like gas adsorption rather than solid adhesion because such heat changes have usually been observed in gas adsorption but not commonly in adhesion between bulk solids. As a result, the adhesion forces between a 2D material and a surface is linearly dependent on temperature. We also present an analytical model based on phonon analysis to elucidate the underlying physics of this intriguing gas-like adhesion of 2D materials.

To gain some insights into the adhesion mechanisms of 2D materials, we first perform two types of atomistic simulations of adiabatically stamping/peeling of a graphene ribbon on/off the (111) surface of a platinum (Pt) substrate. We use the second generation reactive empirical bond order (REBO) [18] to describe the C-C bond interaction in graphene and Lennard-Jones 12-6 potential to describe the interaction between the graphene layer and the substrate (details can be found in the Supplemental Material). We simulate zero-degree stamping and peeling of a graphene ribbon as illustrated in Fig. 1. During the simulations, the graphene ribbon is moved onto or away from the substrate by a constant velocity imposed on the two edges. The Pt substrate is set to be rigid and the temperature change in the graphene layer induced by heat release and extraction is carefully monitored.

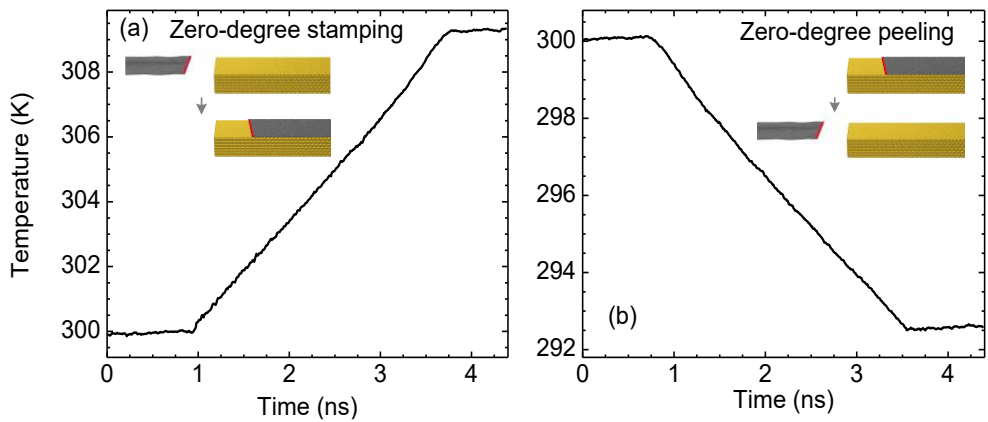

**FIG. 1. The temperature changes during the zero-degree stamping (a) and peeling (b) processes of a graphene ribbon on/off the (111) surface of a Pt substrate. A constant velocity of 10 nm/ns is imposed on the two rigid edges (in red) of the ribbon to move it between the freestanding and the adhered state. The temperature at time *t* is calculated by averaging through a small period of 1 ps around *t*.**

Figures 1(a) and 1(b) show the temperature changes of the graphene ribbon during the stamping and peeling processes, respectively, which take place for 2.85 ns out of a total simulation time of 4.5 ns in both cases. Significant temperature variations show thermodynamic tendencies for heat release during stamping and heat absorption during peeling. To further confirm the heat release/absorption is inherent in the attachment/detachment of the graphene, we simulate the peeling of a graphene ribbon at ninety degrees from a Pt substrate, where a temperature change with same quantity is observed (see Supplemental Material for details). Note that because no other energy form besides mechanical energy is involved here, the heat release in the stamping process is converted from mechanical work, and the heat absorption in a peeling process is converted to mechanical work. The magnitude of the temperature changes in the two cases are quite close, indicating the energy conversions in stamping processes and peeling processes are actually thermodynamically reversible from one state to the other. It also indicates a small friction between graphene ribbon and the Pt substrate, which is consistent with the previous studies on graphene friction [19,20]. The frictionless sliding also explains why the temperature is nearly a constant during the graphene is being totally adhered and sliding on the substrate (the first 0.9 ns in the Fig. 1a and the last 0.9 ns in Fig. 1b) where the friction induced dissipation should reach a maximum due to the maximum contacting area [21].

Although physical adhesion between bulk solid materials may induce heat release in attachment process due to friction, it is impossible to give rise to heat adsorption in detachment. In contrast, gas adsorption on solid surface can release heat and gas desorption form solid surface can absorb heat [22]. In this respect, 2D materials adhesion is indeed gas-like. The similarity of materials nature between gas and 2D materials seems to be responsible for their common thermodynamic adhesion behavior. When gas molecules are adhered on a surface, they form a 2D film that closely conforms to the surface, while 2D materials also can closely conform to a surface due to their extremely flexible membrane structure [23,24]. As a results of the extremely flexible membrane structures, thermal fluctuations are significant in both the 2D materials and the adsorbed, which lead to a fundamental difference in their adhesion behaviors from a bulk solid.

The underlying physics of gas-like adhesion of 2D materials is that the entropy of a free standing 2D layer is larger than that of an adhered one. The mechanism can be attributed to a drop in configurational entropy as the substrate confines the out-of-plane fluctuations of an adhered 2D material layer and reduces its degrees of freedom. Therefore a free standing 2D material layer has a reduced entropy compared to an adhered one. In fact, the entropy change due to external confinement has been extensively studied in many phenomena [25-27]. To confirm the entropy change due to adhesion, we calculated the stamping and peeling of graphene and h-BN under isothermal conditions, following the universal approach used to describe gas adsorptions [22].

The Berendsen thermostat is used to keep the 2D materials at constant temperatures during the attachment and detachment, and the heat exchange between the thermostat and the 2D materials is carefully monitored. Our results show that the heat is released during attachment and absorbed during detachment under isothermal conditions, indicating an entropy reduction in attachment and an entropy increment in detachment. The heat release or absorption induced by the adhesion caused configurational entropy change $\Delta S$ in a reversible process under isothermal conditions can be theoretically expressed as $\Delta E = -T \cdot \Delta S$, $T$ being the temperature. This linear dependence is confirmed by our MD simulations, as shown in Fig. 2a (where we use heat change to represent both heat release and heat absorption because their values are almost the same).

The entropy changes could also generate a phenomenologically entropic force. Thus the adhesion force between a 2D layer and a substrate consists of two parts,

$$F = F_{vdw} + F_e, \qquad (1)$$

where the first term $F_{vdw} = \nabla_x E_{vdw}$ (in which $\nabla_x$ is the differential operator with respect to the path variable $x$, $E_{vdw}$ is the interlayer van der Waals potential) is the van der Waals force, and the second term is the entropic force given by $F_{entr} = -T \cdot \nabla_x S$. At a zero-degree peeling, the adhesion force is indeed the interlayer shear force. To validate Eq. (1), we plot the interlayer shear force of graphene and $h$-BN on Pt substrate from MD simulations as functions of the system temperature in Fig. 2b-c. It is clearly seen that the shear force is linearly dependent on temperature, which is completely caused by the entropic force because the interlayer van der Waals force is a constant.

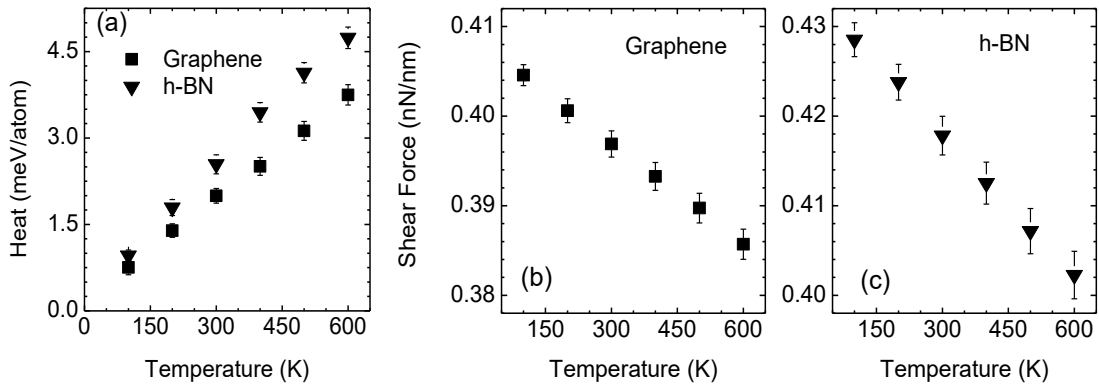

**FIG. 2. (a) Adhesion induced isothermal heat change (heat release in attachment and heat absorption in detachment) versus system temperature. (b) - (c) The shear force (per nanometer in width) versus system temperature.**

In recent years, a number of experimental approaches were developed to determine the adhesion energy of 2D materials on various substrates [29-31]. These approaches are usually based on the force balance between an external loading force and the adhesion force between a 2D layer and a substrate [32]. Typically, at a zero-degree peeling test [30], the critical peeling force, which is equal to the shear force at balance, is measured and used to estimate the adhesion energy. However, as we have shown, the adhesion force consists of not only a conservative van der Waals force but also a entropic force, which would result in a fundamental difference between those experimental measurements and the theoretical predictions $E_{vdw}$ [7,8] derived statically between a 2D layer and a substrate at an equilibrium distance. The ratio of the adhesion force $F_{entr}$ to the van der Waals force $F_{vdw}$ is solely dependent on temperature, and independent of the path variable $x$, indicating this difference is a constant at a given temperature and increases linearly with temperature. In our simulations, at room temperature, this difference is found to be about 3% for the adhesion of graphene on Pt substrate, and 7% for the adhesion between two graphene layers.

Now we intend to elucidate how adhesion causes an entropy change in 2D materials. We analyze the adhesion induced entropy change based on the analysis of normal modes, each of which representing a harmonic mode of vibration of all atoms in a 2D material at a certain frequency. The adhesion is described by a potential well perpendicular to the surface. Under this confining potential, it can be shown that the frequencies of the out-of-plane normal modes are shifted to higher values [28]

$$\omega' = \sqrt{\omega^2 + \kappa/m}, \qquad (2)$$

where $m$ is the atomic mass of the 2D material and $\kappa$ the curvature of the confining potential. A quantum statistical analysis of the normal modes indicates that the increase in frequency results in a decrease in system entropy, as a result of reduced number of microstates (see Supplemental Material for details). Finally, we find that the adhesion induced entropy change of a 2D material can be expressed as

$$\Delta S = \frac{1}{2} k_B \left[ \ln(\frac{\kappa}{\eta}+1) + \sqrt{\frac{\kappa}{\eta}} \cdot \arctan(\sqrt{\frac{\eta}{\kappa}}) \right], \qquad (3)$$

where $k_B$ is Boltzmann's constant, $\eta$ is a constant related to the bending stiffness of the 2D material.

Figure 3 shows the adhesion induced entropy change as a function of potential curvature $\kappa$. It is seen that, form both MD simulations (entropy changes are extracted as $\Delta S = -\Delta E/T$) and Eq. (3), the adhesion induced entropy change logarithmically increases with the increasing potential curvature $\kappa$. For physical adhesion through van der Waals forces, Lennard-Jones 12-6 potential is frequently used to describe the interface interactions. In this case, the potential curvature $\kappa$ depends on the potential well $\varepsilon$ and equilibrium distance $\sigma$, i.e., $\kappa \propto \varepsilon \sigma^{-2}$. Thus the large entropy change on the nickel (Ni) substrate is not caused only by the strong adhesion but also by a small equilibrium distance, which is about 0.21 nm while the equilibrium distances for other substrates are more than 0.3 nm [7,8]. Moreover, a chemical adhesion in general induces a stronger interface interaction (with a larger binding energy and a smaller equilibrium distance) due to the creation of chemical bonds, which can consequently generate a larger entropy change.

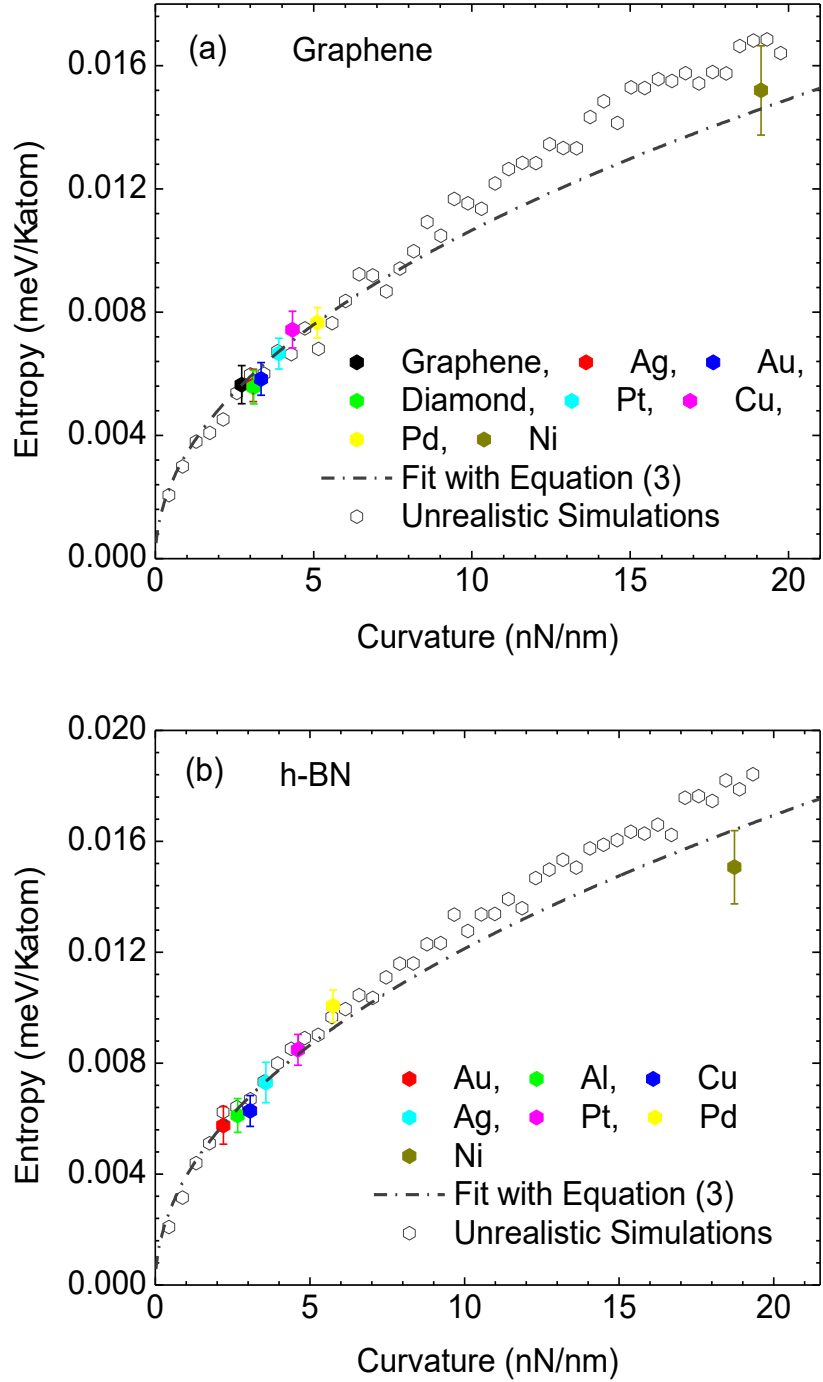

**FIG. 3. (a) Adhesion induced entropy change of graphene on various substrates, including a rigid graphene layer as substrate. The results labelled as "unrealistic simulations" are performed on a Pt substrate with artificial potentials to describe the interaction between atoms of the graphene layer and Pt atoms (see Supplemental S-3 for details). Equation (3) is fitted to MD results with $\eta$ as the fitting parameter. (b) Adhesion induced entropy change of *h*-BN on various substrates.**

Equation (3) also suggests that the adhesion induced entropy change decreases with an increase in the parameter $\eta$, which is positively and almost linearly correlated with the bending stiffness of 2D materials (see Supplemental Material for details). The fitted value of $\eta$ for graphene is about 25% larger than that for *h*-BN (Fig. 3), and the bending stiffness of graphene calculated directly from MD too is 25% larger than that of *h*-BN. Equation (3) also suggests that the gas-like heat release or extraction is vanishingly small in adhesion between bulk materials for which the bending stiffness can be considered infinitely large.

The gas-like adhesion also may raise fundamental issues in many applications of 2D materials. For instance, it has been proposed that using adhesion force as a retracting force could lead to ultrahigh frequency for the oscillator composed of a graphene layer and a substrate [20]. Since the adhesion force is temperature dependent, the frequency of such an oscillator is thus dependent on system temperature. On the other hand, the gas-like adhesion paves a way for developing nanoscale actuators that could convert thermal energy into motions. Designing such devices is yet a challenge due to the absence of an efficient energy conversion mechanism at nanoscale. The energy conversion through 2D material adhesion may overcome this challenge because of its nearly reversible process. In fact, a phenomenon related to this type of energy conversion has already been observed in carbon nanotubes, first in atomistic simulations [15] and then in experiments [33], where the energy conversion is based on a change of the cross-section of the carbon nanotube between its usual circular configuration and a collapsed flat configuration (in which the shell is partially adhered together). It is found that the circular-to-flat collapse and the flat-to-circular restoration are both temperature dependent. While it has been speculated that this phenomenon is governed by changes in system entropy, our present study on the 2D material adhesion provides a clear theoretical foundation to understand such phenomena.

In summary, we demonstrate that adhesion between a 2D material and a substrate is not a typical solid-solid mechanical adhesion but rather a gas-like adhesion. Through attaching/detaching a 2D material onto/from a solid surface, one can convert energy between thermal and mechanical forms. The underlying physics is that the adhesion generate a confinement on 2D materials and cause a drop in their configurational entropy. Because the 2D materials are usually used on substrates, the revealing of their gas-like adhesion provides a more solid foundation for future device integrations.

**Acknowledgements**

Z.G. acknowledges the financial support from China Scholarship Council (File No. 201206890039) during a visit to Brown University. T.C. acknowledges financial support from the NSF (Grants No. 11425209, No. 11172160) of China, and Shanghai Pujiang Program (Grant No. 13PJD016). H.G. acknowledges support from the Center of Mechanics and Materials at Tsinghua University. The authors are grateful to Dr. Teng Zhang and Dr. Hongwei Zhang for informative discussions.## References

[1] A. K. Geim and K. S. Novoselov, Nat. Mater. **6**, 183 (2007).
[2] S. Z. Butler *et al.*, ACS Nano **7**, 2898 (2013).
[3] A. Gupta, T. Sakthivel, and S. Seal, Prog. Mater. Sci. **73**, 44 (2015).
[4] J. Kang, D. Shin, S. Bae, and B. H. Hong, Nanoscale **4**, 5527 (2012).
[5] J. S. Bunch, A. M. van der Zande, S. S. Verbridge, I. W. Frank, D. M. Tanenbaum, J. M. Parpia, H. G. Craighead, and P. L. McEuen, Science **315**, 490 (2007).
[6] J. Rafiee, X. Mi, H. Gullapalli, A. V. Thomas, F. Yavari, Y. F. Shi, P. M. Ajayan, and N. A. Koratkar, Nature Mater. **11**, 217 (2012).
[7] I. Hamada and M. Otani, Phys. Rev. B **82**, 153412 (2010).
[8] M. Bokdam, G. Brocks, M. I. Katsnelson, and P. J. Kelly, Phys. Rev. B **90**, 085415 (2014).
[9] M. Ishigami, J. Chen, W. G. Cullen, M. S. Fuhrer, and E. D. Williams, Nano Lett. **7**, 1643 (2007).
[10] W. Gao and R. Huang, J. Phys. D-Appl. Phys. **44**, 452001 (2011).
[11] K. Yue, W. Gao, R. Huang, and K. M. Liechti, J. Appl. Phys. **112**, 083512 (2012).
[12] N. G. Boddeti, S. P. Koenig, R. Long, J. L. Xiao, J. S. Bunch, and M. L. Dunn, J. Appl. Mech.-Trans. ASME **80**, 040909 (2013).
[13] X. Liu, Q. Li, P. Egberts, and R. W. Carpick, Adv. Mater. Interfaces **1**, 1300053 (2014).
[14] Y. Zhou, Y. Chen, B. Liu, S. Wang, Z. Yang, and M. Hu, Carbon **84**, 263 (2015).
[15] T. Chang and Z. Guo, Nano Lett. **10**, 3490 (2010).
[16] A. Barreiro, R. Rurali, E. R. Hernandez, J. Moser, T. Pichler, L. Forro, and A. Bachtold, Science **320**, 775 (2008).
[17] H. Conley, N. V. Lavrik, D. Prasai, and K. I. Bolotin, Nano Lett. **11**, 4748 (2011).
[18] D. W. Brenner, O. A. Shenderova, J. A. Harrison, S. J. Stuart, B. Ni, and S. B. Sinnott, J. Phys. Condes. Matt. **14**, 783 (2002).
[19] M. Dienwiebel, G. S. Verhoeven, N. Pradeep, J. W. M. Frenken, J. A. Heimberg, and H. W. Zandbergen, Phys. Rev. Lett. **92**, 126101 (2004).
[20] Q. Zheng *et al.*, Phys. Rev. Lett. **100**, 067205 (2008).
[21] Y. Mo, K. T. Turner, and I. Szlufarska, Nature **457**, 1116 (2009).
[22] V. Bolis, *Fundamentals in adsorption at the solid-gas Interface: Concepts and thermodynamics* (Springer, Berlin Heidelberg, 2013), Vol. 1,    p.^pp. 3-50.
[23] C. H. Lui, L. Liu, K. F. Mak, G. W. Flynn, and T. F. Heinz, Nature **462**, 339 (2009).
[24] W. Jung, D. Kim, M. Lee, S. Kim, J. H. Kim, and C. S. Han, Adv. Mater. **26**, 6394 (2014).
[25] W. Helfrich, Z. Naturforschung. **33**, 305 (1978).
[26] G. Costantini and F. Marchesoni, Phys. Rev. Lett. **87**, 114102 (2001).
[27] V. Blickle and C. Bechinger, Nat. Phys. **8**, 143 (2012).
[28] Z. Guo, T. Chang, X. Guo, and H. Gao, J. Mech. Phys. Solids **60**, 1676 (2012).
[29] S. P. Koenig, N. G. Boddeti, M. L. Dunn, and J. S. Bunch, Nat. Nanotech. **6**, 543 (2011).
[30] E. Koren, E. Lortscher, C. Rawlings, A. W. Knoll, and U. Duerig, Science **348**, 679 (2015).
[31] T. Yoon, W. C. Shin, T. Y. Kim, J. H. Mun, T. S. Kim, and B. J. Cho, Nano letters **12**, 1448 (2012).
[32] J. S. Bunch and M. L. Dunn, Solid State Commun. **152**, 1359 (2012).
[33] R. Senga, K. Hirahara, and Y. Nakayama, Appl. Phys. Lett. **100**, 083110, 083110 (2012).
8

# Gas-like adhesion of two-dimensional materials onto solid surfaces: Supplmental materials

Zhengrong Guo[1]*, Tienchong Chang[1,2], Xingming Guo[1], Huajian Gao[3]

[1]*Shanghai Institute of Applied Mathematics and Mechanics, Shanghai University, Shanghai Key Laboratory of Mechanics in Energy Engineering, Shanghai 200072, People's Republic of China*

[2]*State Key Laboratory of Ocean Engineering, School of Naval Architecture, Ocean and Civil Engineering, Shanghai Jiao Tong University, Shanghai 200240, People's Republic of China*

[3]*School of Engineering, Brown University, Providence, Rhode Island 02912, USA*

**Molecular dynamic simulation package and interaction potentials**

The molecular dynamics (MD) simulations were performed based on the Large-scale Atomic/Molecular Massively Parallel Simulator code (LAMMPS) [1] and the Verlet algorithm with a time step of 1 fs . The second generation reactive empirical bond order (REBO) [2] was used to describe the C-C bond in graphene, and the Tersoff potential[3] to describe the B-N bond in *h*-BN. These potentials are being widely used to investigate the mechanical and thermal properties of 2D materials. A Lennard-Jones (LJ) 12-6 potential with cutoff distance of 1 nm was used to describe the graphene/substrate and *h*-BN/substrate interactions [4]. The values of LJ parameters $\varepsilon$ and $\sigma$, obtained from fitting with DFT calculations [4-6], are listed in Table 1. Note that, although the interactions between graphene and some metal substrates may involve hybridizations [5], it has been shown that the LJ potential can capture the adhesive energy as a function of the inter-surface distance quite accurately. In addition, there are two types of atoms in the *h*-BN layer, with uncertain parameter values for the LJ potential. In current simulations, the same parameters were used for both types of atoms in *h*-BN. This approximation may have some quantitative effects on the simulation results, but should not change our main conclusions.

**Attachment and detachment of the graphene ribbon on/off a substrate**

The size of the Pt substrate used in the first two simulations is 41 nm (length) × 9.7 nm (width) × 4 nm (height), while the size of graphene ribbon is 28.5 nm (length) × 9.7 nm (width). Periodical boundary condition is applied in the width direction. We first performed the simulation of peeling the graphene ribbon from the Pt substrate (FIG. 1c), and took its final configuration as the initial configuration to perform the stamping simulation (FIG. 1b).

In the first simulation (FIG. 1c), a graphene ribbon is initially placed on the (111) surface of the Pt substrate with its edge at 0.95 nm from the left sidewall of the substrate. Before we perform the simulations using an NVE ensemble, we first relax the graphene for 200 ps at 300K using an NPT ensemble to remove any possible pressure in the width direction. Because the relaxation is performed only for graphene ribbon, the substrate may be slightly stretched or compressed in the width direction. The temperature of graphene at time *t* is obtained by averaging the instantaneous temperatures through a small time period of 0.9 ps around the time *t*.

To rule out the possibility that the temperature change relies on the path by which the graphene ribbon is peeled off the substrate, we conduct another peeling simulation in which the graphene ribbon is peeled from the Pt substrate at 90 degree angle with a constant velocity of 10 nm/ns (FIG. S1a). The temperature change during the peeling process is shown in FIG. S1b.

**Table 1.** Parameters for the 12-6 Lennard-Jones potential. The parameters were fitted from DFT calculations for Metal/C interaction [3] and Metal/BN [4]. The $C^a$/C is for graphene/graphene interaction, and $C^b$/C for graphene/diamond interaction [5], and $\kappa$ is the curvature at the bottom well of the LJ potential.

| Graphene | Ni/C | Cu/C | Au/C | Pt/C | Ag/C | Pd/C | $C^a$/C | $C^b$/C |
|---|---|---|---|---|---|---|---|---|
| $\varepsilon$ (meV) | 39.50 | 10.40 | 8.60 | 9.09 | 7.95 | 12.0 | 2.97 | 3.15 |
| $\sigma$ (nm) | 0.210 | 0.298 | 0.330 | 0.330 | 0.326 | 0.295 | 0.3407 | 0.342 |
| $\kappa$ (N/m) | 19.136 | 4.334 | 3.396 | 3.903 | 3.094 | 5.120 | 2.724 | 3.107 |
| **h-BN** | Ni/BN | Cu/BN | Au/BN | Pt/BN | Ag/BN | Pd/BN | Al/BN | |
| $\varepsilon$ (meV) | 31.31 | 7.50 | 5.53 | 10.5 | 8.91 | 13.12 | 6.15 | |
| $\sigma$ (nm) | 0.206 | 0.332 | 0.328 | 0.331 | 0.341 | 0.305 | 0.372 | |
| $\kappa$ (N/m) | 18.727 | 3.069 | 2.20 | 4.613 | 3.576 | 5.751 | 2.668 | |

**Statistics of the peeling force and adhesion heat (entropy)**

We calculated the peeling force and the adhesion heat in the same way described in section S-1, except that an NVT ensemble is used to keep the system at constant temperatures. The calculated peeling force and adhesion induced heat fluctuate with time, and both are Gaussian-distributed, as shown in FIG. S2. By fitting their distributions to Gaussian functions, we obtained the average values for the peeling force and adhesion induced heat. The existence of friction may slightly bias the adhesion heat and peeling force we measured. The measured heat in a peeling process is $\Delta E^s_{\text{measure}} = \Delta E - \Delta E_{\text{Friction}}$, while the measured heat in an stamping process is $\Delta E^a_{\text{measure}} = \Delta E + \Delta E_{\text{Friction}}$. Thus the adhesive heat without frictional bias can be obtained by $\Delta E = (\Delta E^a_{\text{measure}} + \Delta E^s_{\text{measure}})/2$. The peeling forces is obtained in the same way.

In our simulations, all atoms in the substrates were fixed to eliminate fluctuations of the calculated results. To assure this setting has no significant influence on our conclusion, we performed simulations of a graphene ribbon supported by a Pt substrate with its atoms freely vibrated. In this case, we found that the adhesive heat is difficult to be measured because of the relatively very large fluctuations due to the large amount of the substrate atoms. Fortunately, the peeling force on the graphene ribbon can still be calculated as shown in FIG. S3. However, the entropic part of the peeling force in this case is about 50% larger than that on the rigid substrate. This may be attributed to two possible mechanisms. First, the thermal vibration of substrate can induce fluctuations of the out-of-plane motion of 2D layers, which may increase the adhesion entropy. Second, the quasi-2D structure of the substrate experiences significant out-of-plane thermal vibrations, which could enhance the adhesion caused entropy too. Moreover, the substrate is stacked up by layers of atoms arranged in (111) planes. Although they are chemically bonded together, their out-of-plane vibration still exist, especially for the top surface layer. However, a theoretical estimate of the contribution of these mechanisms is difficult and material-dependent, thus the present study focus only on the adhesion entropy contributed by the 2D layers.

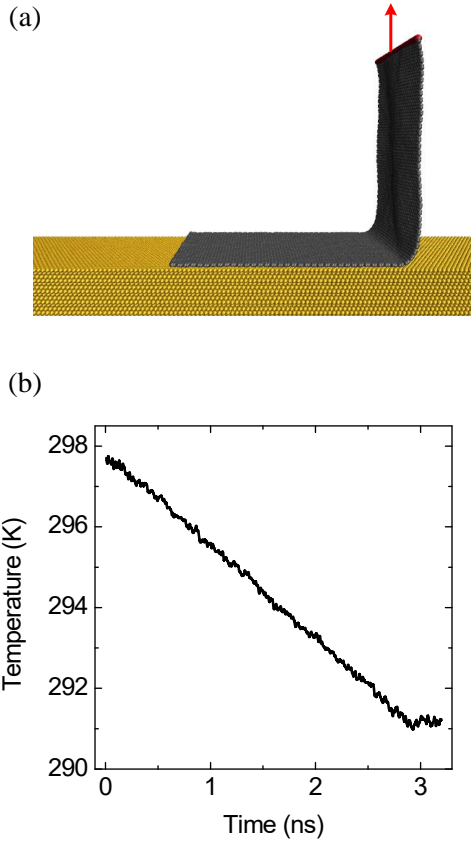

FIG. S1. Ninety-degree peeling of a graphene ribbon from a Pt substrate. (a), A snapshot of the configuration, and (b), The temperature change of the graphene ribbon during the peeling process.

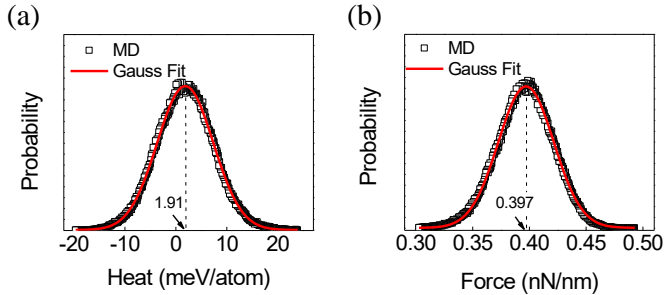

FIG. S2. Distribution of instantaneous values for the peeling force (a) and adhesion induced heat change (b) for the zero-degree peeling of a graphene ribbon off a Pt substrate at 300K.

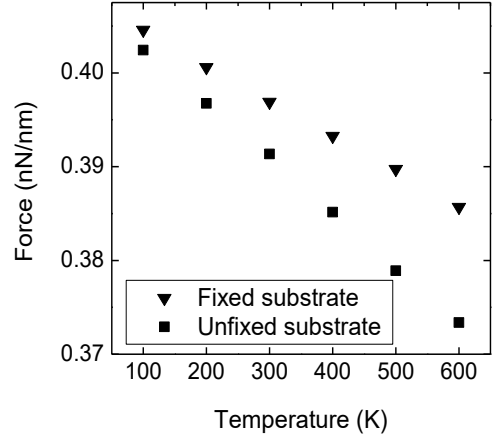

FIG. S3. Comparison of peeling force on the rigid and flexible substrates.

**Unrealistic simulations**

The unrealistic simulations are performed on a rigid substrate have the same crystalline structure as Pt, with the artificial potential parameters for layer/substrate interaction: a constant equilibrium distance of 0.33 nm and a varying well-depth from 1 meV to 50 meV, for yielding different curvature $\kappa$.

**Derivation on the adhesion induced entropy reduction**

Here we consider the simplest situation of a 2D crystal with $N$ lattices, and each lattice cell contains only one atom with mass $m$. The expression of per atom entropy of this system can be given by

$$S = \frac{k_B}{N} \ln \Omega \tag{S1}$$

where $k_B$ is Boltzmann constant, $\Omega$ is the total number of the microstates. Only those microstates related to out-of-plane normal modes (phonon) are considered. The number of microstates for a normal mode with frequency $\omega_i$ is given by

$$\Omega(\omega_i) = \frac{1}{e^{\hbar\omega_i/k_B T} - 1} \approx \frac{k_B T}{\hbar \omega_i} \tag{S2}$$

where $T$ is the system temperature, $\hbar$ is the Planck constant. The approximation in equation (S2) breakdown only when the temperature becomes extremely low. The total number of microstates of all the out-of-plane normal modes is

$$\Omega = \prod_{i=1}^{N}(k_B T / \hbar \omega_i) \quad (S3)$$

If harmonic approximation is adopted for describing the influence of adhesion on the normal vibration, the frequency of the out-of-plane normal modes is shifted to

$$\omega' = \sqrt{\omega + \kappa/m} \quad (S4)$$

where $\kappa$ is the force constant of the harmonic potential (for a general potential, it is the curvature of the potential at bottom). With those equations, the per atom adhesion entropy between freestanding state and adhered state can be written as

$$S - S^\dagger = \frac{1}{N} k_B \sum_{i=1}^{N} \ln\sqrt{1 + \kappa/m\omega_i^2} \quad (S5)$$

The equation (S5) can be recast in an integral form

$$S - S^\dagger = \frac{1}{N} k_B \int_0^{\omega_{max}} \rho(\omega) \cdot \ln\sqrt{1 + \kappa/m\omega^2}\, d\omega \quad (S6)$$

where $\rho(\omega)$ is the distribution of the density of normal modes (DOS). It can be seen from Equation (S5) that the entropy change is mainly dependent on those modes with low frequency. In theoretical studies, a distribution of $\rho(\omega) = c\omega^\tau$ (where $c$ and $\tau$ are constants) is usually used for those normal modes. For 2D materials, the distribution of out-of-plane modes is $\rho(\omega) = c$. A maximal frequency $\omega_{max}$ can be defined for the existence range of frequency. Its relation with the total number of modes is given by $N = \rho(\omega)\omega_{max}$. In fact, the maximum frequency $\omega_{max}$ is a material-dependent variable, which is linearly dependent on the dihedral force constant, while the bending rigid of 2D materials $G$ is also linear dependent on dihedral constant. Thus we expect $\omega^2_{max} \propto G$.

Finally, the adhesive entropy is given by

$$S - S^\dagger = \frac{1}{2} k_B \left\{ \ln(\frac{\kappa}{\eta} + 1) + \sqrt{\frac{\kappa}{\eta}} \cdot \text{Arctan}(\sqrt{\frac{\eta}{\kappa}}) \right\} \quad (S6)$$

where $\eta = m\omega_{max}^2$ is also a material-dependent constant.

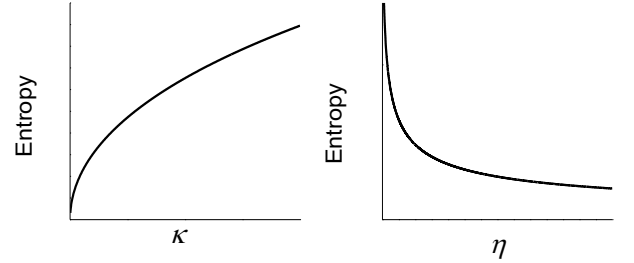

**FIG. S4.** Adhesion induced entropy change versus curvature $\kappa$ and material-dependent parameter $\eta$.

The derivation above can be easily extended to the situation that each lattice cell contains two atoms. For graphene, because the two atoms in each unit cell have the same atoms, thus the equations from (S1) to (S6) have exactly the same forms. For h-BN, the only difference is the mass and it is given by $m_{BN} = 2m_B m_N/(m_B + m_N)$, where the $m_B$ and $m_N$ are the mass of boron atom and nitride atom, respectively.